# Advancing Precision Oncology Through Modeling of Longitudinal and Multimodal Data

Luoting Zhuang, *Graduate Student Member, IEEE*, Stephen H. Park, Steven J. Skates, Ashley E. Prosper, Denise R. Aberle, William Hsu, *Senior Member, IEEE*

***Abstract*—Cancer evolves continuously over time through a complex interplay of genetic, epigenetic, microenvironmental, and phenotypic changes. This dynamic behavior drives uncontrolled cell growth, metastasis, immune evasion, and therapy resistance, posing challenges for effective monitoring and treatment. However, today's data-driven research in oncology has primarily focused on cross-sectional analysis using data from a single modality, limiting the ability to fully characterize and interpret the disease's dynamic heterogeneity. Advances in multiscale data collection and computational methods now enable the discovery of longitudinal multimodal biomarkers for precision oncology. Longitudinal data reveal patterns of disease progression and treatment response that are not evident from single-timepoint data, enabling timely abnormality detection and dynamic treatment adaptation. Multimodal data integration offers complementary information from diverse sources for more precise risk assessment and targeting of cancer therapy. In this review, we survey methods of longitudinal and multimodal modeling, highlighting their synergy in providing multifaceted insights for personalized care tailored to the unique characteristics of a patient's cancer. We summarize the current challenges and future directions of longitudinal multimodal analysis in advancing precision oncology.**

## I. Introduction

CANCER, a complex and heterogeneous disease, continues to be a leading cause of morbidity and mortality worldwide [1]. The disease originates from genetic mutations, leading to abnormal cells that continue to proliferate uncontrollably, invade surrounding tissues, and metastasize to distant sites. Cancer's ongoing evolution also allows it to evade the immune system, develop resistance to therapies, and recur at the original site, which poses significant challenges for effective monitoring and treatment [2]. Hence, advanced analytical methods are necessary to capture and interpret its dynamic and multifaceted behavior. Longitudinal and multimodal analysis can be pivotal in this regard, offering frameworks for integrating diverse data collected repeatedly throughout a patient's healthcare journey to discover longitudinal multimodal biomarkers.

Longitudinal analysis involves modeling data accumulated over time, which enables the examination of temporal patterns in cancer biomarkers. This is in contrast to cross-sectional analysis, which only provides a snapshot of the disease at a single timepoint. Numerous studies have highlighted the benefits of longitudinal analysis over cross-sectional analysis in the medical domain [3], [4], [5], [6], [7], [8], [9], [10], [11], [12]. For instance, in cancer screening, many studies demonstrate that longitudinal modeling of serial protein biomarkers in blood outperforms the single threshold method [4], [5], [7], [9]. Moreover, changes in biomarkers before and after initial therapy or from one disease course to the next are often associated with cancer prognoses [10], [11].

Statistical methods have conventionally dominated modeling longitudinal data to facilitate clinical decision-making. While statistical methods focus on population inference of a sample dataset to understand the data-generating process, artificial intelligence (AI)-driven approaches concentrate on prediction optimization by finding generalizable patterns in the data [13]. With the exponential growth of data collection and the incorporation of unstructured data from various sources, AI methods demonstrate considerable promise in the domain [14], [15], [16]. Particularly, the emergence of machine learning (ML) and deep learning (DL) techniques has allowed AI to discern intricate temporal patterns, detect subtle correlations, and offer enhanced predictive power in cancer diagnosis and prognosis. Despite these advancements, DL models still face several challenges, including irregularly sampled data, high computational burden, and lack of interpretability. Therefore, the choice of the appropriate model requires careful consideration of numerous factors, such as data volume, feature complexity, temporal regularity, and clinical task.

The majority of longitudinal analysis in cancer research thus far has involved the modeling of single modalities. For instance, as previously mentioned, many cancer screening studies model the changes in blood biomarkers to assess the likelihood of having or developing undetected cancer. However, this approach may not fully exploit the data available from other modalities, such as family history of cancer or screening imaging exams. In fact, a variety of data are collected as part of routine clinical care, including clinical records, molecular profiles, medical imaging, and histology [17], [18], [19]. As

Manuscript submitted 16 December 2024; revised 24 April 2025; accepted 1 June 2025. This work was supported by NIH/National Cancer Institute U2C CA271898, U01 CA233370, and the V Foundation. (Luoting Zhuang and Stephen H. Park are co-first authors.) (Corresponding author: William Hsu.)

Luoting Zhuang, Stephen H. Park, Ashley E. Prosper, Denise R. Aberle, and William Hsu are with Medical & Imaging Informatics, Department of Radiological Sciences, David Geffen School of Medicine at UCLA, Los Angeles, CA 90024 USA (e-mail: luotingzhuang@g.ucla.edu; sparkheejae@g.ucla.edu; aprosper@mednet.ucla.edu; daberle@mednet.ucla.edu; whsu@mednet.ucla.edu)

Steven J. Skates is with MGH Biostatistics, Massachusetts General Hospital, Somerville, MA 02145 USA (e-mail: sskates@mgb.org)

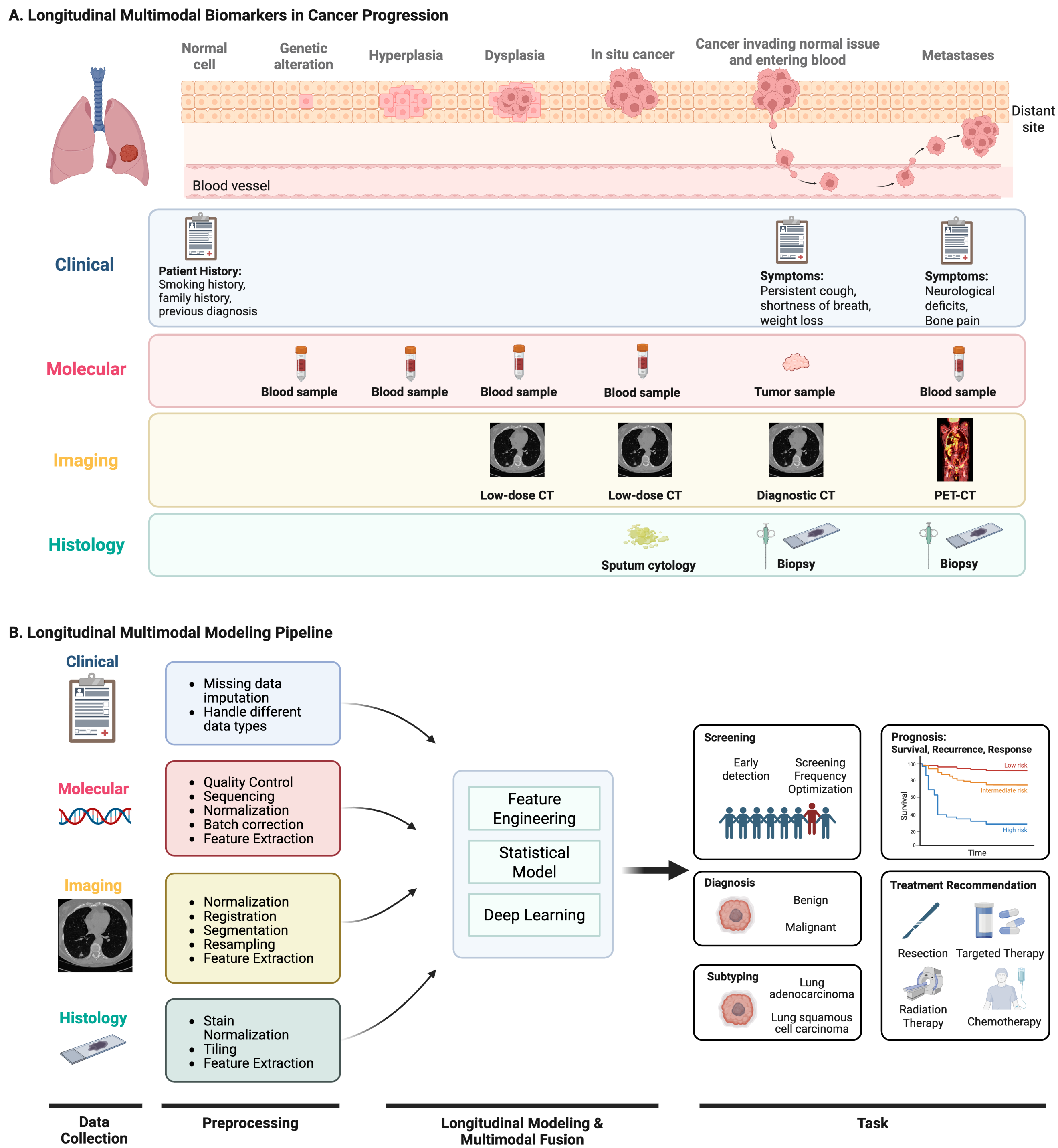

**Fig 1. Longitudinal multimodal data and modeling pipeline.** Figure **(A)** represents an example of longitudinal multimodal data that can serve as biomarkers throughout different stages of lung cancer progression. Each data modality provides a unique perspective into the complex disease process. **Clinical** records like patient history can offer insights into cancer risk even at the normal cell stage. Symptoms such as persistent cough, weight loss, and bone pain can signal disease progression at later stages. **Molecular** profiles from blood samples collected at various stages of cancer progression, from genetic alteration to metastases, provide crucial information about the underlying biological changes in the tumor. Notably, molecular data obtained from blood samples can serve as early cancer biomarkers, even when other modalities like imaging or histology might not yet detect any signal. **Medical imaging** becomes particularly valuable when phenotypic changes occur, such as the formation of pre-cancerous lung nodules due to dysplasia. Low-dose CT can be used in the early stages for assessing the location, size, shape, and texture of nodules, whereas whole-body PET-CT scans can help identify and analyze clusters of tumor cells that have metastasized. **Histology** data from sputum cytology or biopsy are often used for definitive cancer diagnosis, revealing detailed information on tissue architecture and cellular morphology. Figure **(B)** shows the general pipeline of longitudinal multimodal data analysis. The workflow consists of longitudinal multimodal data collection, modality-specific preprocessing, modeling, and clinical application. Created in BioRender. Hsu, W. (2025) https://BioRender.com/r93v294.

illustrated in Fig. 1A, each modality contributes unique and potentially valuable signals throughout different stages of disease progression. Hence, multimodal analysis may further enhance the capacity of longitudinal models by harnessing the strengths of different modalities to characterize tumors comprehensively [20], [21], [22].

In this review, we explore the characteristics and acquisition methods of various data modalities used in longitudinal analysis, methods for extracting predictive temporal features, and approaches for representing and modeling longitudinal multimodal data. We also identify significant gaps and challenges in the field, along with potential solutions.

## II. Longitudinal Multimodal Data

Throughout a patient's journey in healthcare, diverse data are obtained from various modalities to understand the patient's condition and inform management (Fig. 1A). Many of these data are repeatedly collected over time, providing opportunities for longitudinal multimodal analysis. In this review, we primarily focus on four modalities commonly acquired in oncology: clinical, molecular, imaging, and histology. Each modality is unique in nature and utility yet collectively important for comprehensive patient assessment (Fig. 1B). Here, we explore the characteristics of these modalities, methods of acquisition, and potential value in longitudinal analysis. We also provide a table of publicly available cancer datasets featuring longitudinal multimodal data (Supplemental Table I). To note, our definitions of these categories are not universal but rather serve as a framework to guide our discussion on each modality's specific uses and contributions in the context of longitudinal multimodal analysis.

### *A. Clinical*

Clinical data represent information gathered from patients during their interaction with healthcare systems. This data category includes a wide variety of features related to demographics, diagnosis, medication, and treatment. The acquisition of clinical data is largely facilitated by electronic health records (EHRs), which allow for the systematic collection and storage of patient information across different points of care [23]. These records not only improve the efficiency of healthcare systems but also offer benefits to research by providing structured datasets that are relatively easy to query and analyze.

Clinical data are suitable for longitudinal analysis because many metrics are collected continually throughout the patient's clinical workflow. In oncology, routine clinical and laboratory data collection can provide insights into cancer risk even at the normal cell stage. This presents opportunities for early monitoring, such as participation in screenings and incidental cancer detection [24], [25], [26]. Additionally, clinical data alongside symptoms reported by patients can suggest disease progression, adverse events related to cancer treatments [27], [28], metastases [29], and recurrence [30].

However, variations in clinical data collection across time can introduce significant bias into the analysis. For example, the evolution of diagnostic codes over time can change how conditions are categorized. In addition, inconsistencies among healthcare providers in reporting and documenting patients' information and differences in instruments used for measurements can lead to systematic variability in the data.

### *B. Molecular*

Molecular data encompasses a broad range of data types derived from molecules. These include genomic, epigenomic, transcriptomic, and proteomic data. The source of molecules can either be solid tissues or biofluids such as blood, serum, and saliva. In oncology, interest has been shifting more and more toward the sampling and molecular analysis of tumor-derived analytes in accessible biofluids, typically referred to as liquid biopsy (LB) [31], [32]. Common analytes in LB include cell-free DNA (cfDNA), circulating tumor cells (CTCs), and proteins. For cfDNA, technologies such as Polymerase Chain Reaction (PCR) and Next Generation Sequencing (NGS) are used along with bioinformatics algorithms to obtain data related to the abundance of cfDNA, somatic mutations, copy number alterations, DNA methylation, or DNA fragmentations. Methods like the CellSearch [33] and microfluidic devices are employed to isolate and enumerate CTCs. As for proteins, immunoassays are typically used to measure their concentrations.

LB has the potential to provide significant value in longitudinal studies because its minimally invasive collection allows relatively easy sampling of serial biospecimens. For instance, cfDNA can be extracted from biospecimens collected during routine peripheral blood tests. Research has shown that longitudinal LB can help detect cancer early, monitor cancer evolution, identify the emergence of treatment resistance, and assess treatment efficacy [34], [35]. However, there are significant challenges that may hinder the extensive collection of sequential data from LB. Extracting low-concentration analytes from biofluids and processing them requires advanced technology and specialized expertise, as well as thorough quality control. Moreover, a vast amount of longitudinal biospecimens would need to be stored and preserved at appropriate conditions through well-maintained biobanking systems to fully characterize the range of cancer biomarker trajectories, which can be a major hurdle in resource-limited settings.

### *C. Medical Imaging*

Medical imaging visualizes the internal structures and functions of the body for clinical assessment. There is a range of modalities in medical imaging, including planar radiography, Ultrasound, Mammography, Magnetic Resonance Imaging (MRI), Computed Tomography (CT), and Positron Emission Tomography (PET) [36]. These imaging modalities produce two-dimensional (2D) or three-dimensional (3D) scans that can either be used directly for analysis or further processed to extract radiomics or DL-based "deep" features from a region of interest (ROI). Although less prevalent than 2D and 3D scans, four-dimensional imaging offers a dynamic representation of structures over a short period of time to better visualize the moving organs and blood flows [37].

Medical images are typically acquired from the same subject multiple times during their healthcare journey. For instance, multiple imaging screenings are performed to detect and monitor early signs of cancer [38]. Diagnostic imaging is used to confirm the presence of cancer, and additional imaging may be performed throughout the treatment process to assess the patient's response to therapy [39], [40]. However, medical imaging data are collected with relatively low frequency to minimize cumulative radiation dose, reduce the adverse effects on patients' health, and avoid substantial financial cost. Moreover, although an increasing number of patients have longitudinal scans in clinical practice, their data are typically not publicly accessible. Publicly available datasets often come from clinical trials, which vary in structure and may not fully reflect real-world settings. Consequently, analyses conducted on such data may yield models that do not directly translate to the clinic. In addition, limited funding for clinical studies can restrict the number of medical imaging procedures due to their high cost and the need for specialized equipment and trained technicians.

### *D. Histology*

Histology data refer to images that are obtained from the microscopic examination of stained tissue sections, revealing cellular components and structures. The process begins with the extraction of tissue samples from a suspected tumor requiring either invasive biopsy or surgical resection. Pathologists then review the morphology and structure of tissue samples, acquiring information such as mitotic activity, enlarged blood vessels, necrosis, and immune cell infiltration, to confirm the diagnosis and determine the grading of cancer. These morphological features are also significantly correlated with the survival outcomes of cancer patients [41]. With the Food and Drug Administration (FDA) approval of Whole Slide Imaging (WSI), histology images are digitized to be shared and analyzed electronically [42].

Histopathology results are often considered the gold standard for diagnosis in oncology. However, when core needle biopsies fail to fully capture the tumor, and given the inherent 2D nature of histological slides, there is a risk of misdiagnosis and interrater variability. Moreover, collecting longitudinal histology images is highly challenging due to several factors. Biopsies are invasive procedures that can cause discomfort and present potential risks to patients. As a result, the frequency of these procedures is often kept to a minimum. Secondly, in certain circumstances, complete resection of the tumor may be performed, thereby precluding the need for a follow-up biopsy on the same tumor. Despite these challenges, there can be instances where a metastasized or recurrent tumor is biopsied again, and such a subsequent biopsy can reveal potential histologic changes compared to the original tumor.

## III. Longitudinal Modeling

Analyzing longitudinal medical data can facilitate understanding disease progression, monitoring patients' outcomes, and discovering longitudinal biomarkers for clinical decision-making. A simple approach that has been widely used is to engineer features that capture the dynamic nature of longitudinal data. Numerous statistical models have been developed and employed to analyze trends in repeated measurements. More recently, DL has shown promise in uncovering complex temporal dependencies and nonlinear relationships within longitudinal medical data. In this section, we delve into the details of each method, exploring its applications in analyzing temporal data throughout various medical domains. We also outline the strengths and limitations of each method (Table I). Relevant formulas are provided in Supplemental Table II. It should be noted that the methods we introduce in this section are not solely derived from cancer-related studies, as methods that are utilized in other disease domains have potential for application in oncology.

### *A. Feature engineering*

Feature engineering involves the manipulation of raw data to extract more useful features that can be used for downstream tasks. Once extracted, the features are usually fed into task-specific statistical, ML, or DL models that accept non-sequential data as input. Common downstream models are Cox regression (CR), logistic regression (LR), random forest (RF), support vector machine (SVM), K-nearest neighbors (KNN), multi-layer perceptron (MLP), and convolutional neural network (CNN). While some models inherently include feature engineering as part of their design, these models rely on additional techniques to extract, transform, or construct features that effectively capture the underlying patterns in the data. In longitudinal analysis, it is important that these techniques extract the temporal information provided by the serial data points. The methods we present in this section involve handcrafting techniques or non-DL algorithms that aim to achieve this.

**Concatenation.** One feature engineering method is concatenation, where the data from each timepoint are appended together into the same feature space (Fig. 2A). Serial feature vectors that are generated from any modality can simply be attached together to be fed into ML or DL models [25], [43], [44]. When dealing with higher dimensional data, such as imaging, 3D volumes of images from multiple timepoints can be stacked together in the channel dimension before being input into DL models such as CNNs [45]. Alternatively, before concatenation, one-dimensional (1D) radiomics or deep features can be extracted from the images [46]. Since concatenation does not provide the model with any information on the duration or change between temporal observations, one way to include this information is to concatenate an additional feature that encodes time, such as the time difference between datapoints [47].

**Extraction of predefined features.** Another intuitive method of encoding temporal information is characterizing the change in features between timepoints based on predefined functions (Fig. 2B). Generally, two types of features that describe change can be engineered: numerical and categorical. Features that indicate numerical change are those that explicitly quantify the change in datapoints over time. Examples include the absolute difference, relative difference, and rate of difference [3], [5], [11], [24], [26], [40], [43], [48], [49], [50]. In contrast, categorical features assign labels to specific patterns of change in datapoints. Common categories include "increase", "decrease", "stable", or "change" [10], [51], [52], [53]. These numerical and categorical change features are subsequently used as inputs to statistical or ML models for risk analysis and classification tasks [54], [55], [56].

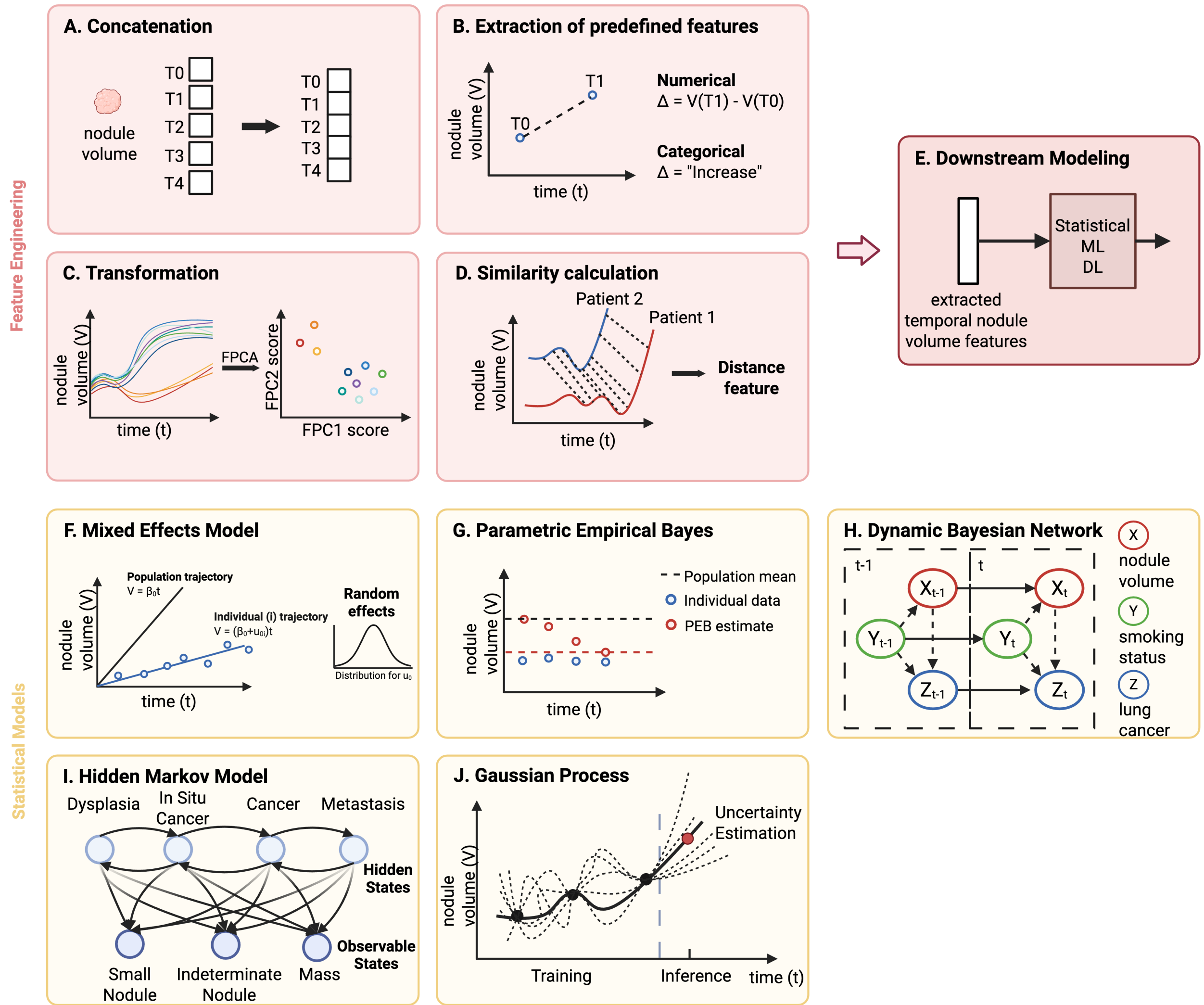

**Fig. 2. Feature engineering and statistical models.** All models are illustrated in the context of lung cancer, with lung nodule volume as the main feature of interest. Figure (A-D) represent four different **feature engineering** techniques to extract temporal features. **(A)** Features from each time point (T0-T4) are concatenated together. **(B)** Change in features between timepoints T1 and T0 are represented numerically or categorically based on predefined functions. **(C)** The longitudinal trajectories are transformed to functional principal component scores using functional principal component analysis, an algorithmic transformation. **(D)** The similarity between two trajectories is computed using a non-linear alignment algorithm (e.g., dynamic time warping). **(E)** After feature extraction, the temporal features are fed into statistical, machine learning, or deep learning models to perform various clinical tasks. Figure F-J show five **statistical models** commonly used for modeling sequential data. **(F)** A linear mixed effects model parametrizes the individual-specific linear trajectory, composed of fixed effects (β) and random effects (u). The random effects represent the deviation from the population trajectory and are often parametrized with normal distributions. **(G)** Parametric empirical Bayes (PEB) is applied to estimate the individual-specific parameter. The parameter estimate starts from the population mean and iteratively updates based on the individual's data at each timepoint, resulting in an individual-specific PEB estimate of the parameter. **(H)** A dynamic Bayesian network depicts sequences of features and their probabilistic interactions over time in a graphical structure. The state of a feature at timepoint (t) is dependent on other features at the same timepoint and those at the preceding timepoint (t-1). **(I)** A hidden Markov model represents the temporal sequence of observations, which rely on temporal hidden states. The edges represent the transition probabilities between states and the emission probabilities, indicating the likelihood of an observable state given the hidden state. **(J)** Gaussian process is used to perform inference and capture uncertainty in the temporal data by representing the possible trajectory functions between observations as continuous functions, with a joint Gaussian distribution over function values at different timepoints. Created in BioRender. Hsu, W. (2025) https://BioRender.com/w23g935.

**Transformation.** Applying algorithmic transformations is also a common method of feature engineering. These transformations are usually automated and applied to functional data. One example of a transformation is functional principal component analysis (FPCA), shown in Fig. 2C. FPCA treats each individual's feature over time as a function and aims to capture the largest amount of variation in all the functions using the least amount of functional principal components (FPCs). FPC scores are obtained by projecting each individual's trajectory onto the FPCs and can subsequently be used as features in downstream models for various tasks [57], [58], [59]. For instance, FPC scores derived from molecular biomarker trajectories have been input into CR models for the prediction of survival in cancer patients [60]. Discrete wavelet transform (DWT) is another example of a transformation, widely used in the neurogenerative disease domain [61], [62], [63]. It decomposes a time series into both time and frequency domains, providing localized insights into when specific frequencies occur. The result of a DWT is a set of wavelet coefficients that represent the time series at different frequencies and times, capturing both gradual and rapid changes. The coefficients can then be used directly or further processed (e.g., mean, variance) as features for downstream models. Moreover, an example of a transformation that induces dimension expansion is the random convolutional kernel transform (ROCKET). ROCKET uses numerous random 1D convolutional kernels to transform a time series and generate a large vector of features for use in downstream tasks. This method has been applied to transform longitudinal clinical data, such as vital signs and heart rates, to perform classification tasks for COVID-19 and fetal cardiotocography, respectively [64], [65].

**Similarity calculation.** Distance-based similarity features can be engineered to classify trajectories. Dynamic time warping (DTW) is a technique widely employed to quantify the similarity between time series (Fig. 2D). This method identifies the most optimal alignment between two sequences and calculates the distance between them. DTW distance can then be used as the similarity metric for unsupervised clustering models or supervised classification models such as KNN [66], [67]. Alternatively, pairwise distances between an individual's trajectory and all other trajectories can be stacked into a feature vector and input into downstream models like SVM [68], [69]. Shapelet transformation is another method of engineering similarity features. It involves selecting shapelets, or subsequences, from a training set of trajectories and computing the distances between a test trajectory and each of the selected shapelets. This results in a vector of distances for each trajectory, which can then be fed into various classifiers. Shapelet transformation has been widely employed for analyzing longitudinal clinical and molecular data in medical domains outside of oncology [70], [71], [72].

### *B. Statistical models*

Statistical models are designed to represent the data-generating process by describing the relationships between outcome variables and predictors using equations and distributional assumptions. When modeling longitudinal data, statistical models use various frameworks to capture the relationships between timepoints, predictors, and response variables. A wide range of statistical models has been explored to analyze longitudinal medical data.

**Mixed effects model.** A mixed effects model (MEM) is a statistical model that can incorporate both fixed (population) and random (individual) effects. Fixed effects are parameters that are constant across all individuals in the population, whereas random effects represent parameters that can vary by individual (Fig. 2F). In the context of modeling longitudinal data, the population-average trajectory of the feature of interest can be modeled with fixed effects, while the deviations of each individual's feature from the general trajectory can be captured by random effects. Although there are many variations of MEMs, they can be generally categorized as linear mixed effects models (LMEMs) or nonlinear mixed effects models (NLMEMs). LMEMs, the most general form of MEMs, assume a linear relationship between the response feature and the predictor. When performing longitudinal analysis, the predictor is typically defined as time, parametric functions of time, or other time-varying variables. The coefficients, composed of fixed and random effects, indicate the intercept and slope. On the other hand, NLMEMs handle nonlinear relationships involving more complex mechanisms, such as ordinary differential equations (ODEs), to describe the dynamics of longitudinal data [73], [74], [75]. NLMEMs, while more computationally demanding than LMEMs, provide the flexibility needed to effectively model complex, nonlinear processes.

MEMs have been used within the framework of mixture models. Mixture models assume that the observed data are generated from a mixture of different distributions, each representing a latent unobserved subgroup. Mixture models are particularly useful in longitudinal analysis because they can model a mixture of two or more subgroups, each with similar feature trajectories over time. Within each subgroup, class-specific MEMs describe the longitudinal behavior of observed features [76]. For example, a widely studied cancer screening algorithm uses a hierarchical mixture change-point model to characterize the behavior of molecular biomarker trajectories [6], [77], [78]. The model is structured under the assumption that patients without cancer have a stable biomarker trajectory, whereas patients with cancer have a mixture of two types of trajectories depending on whether cancer sheds the marker or not.

MEMs can also be jointly modeled with risk models to perform prognostic tasks. Joint modeling facilitates analysis of relationships between longitudinal processes and the hazard of outcomes by employing shared parameters that link the two underlying models. For instance, a time-dependent feature derived by a MEM can be used as an additional covariate in a Cox proportional hazards model. Typically, a parameter that incorporates the unobserved random effects from MEMs is used as the shared parameter, with the assumption that the individual deviations from the population-average trajectory influence the risk of the event. In oncology, joint models have been frequently used to model longitudinal protein level data for cancer outcome analysis and have outperformed models that rely only on baseline covariates [74], [76], [79], [80], [81].

**Parametric empirical Bayes.** Parametric empirical Bayes (PEB) is a statistical procedure used to estimate individual-specific parameters, integrating the principles of Bayesian statistics with empirical data [82]. In standard Bayesian analysis, the prior distributions for unknown parameters are specified before observing any data. PEB, in contrast, uses the observed data across the population to estimate the parameters of a prior distribution. Then, for each individual, a Bayes estimator is used to iteratively update the prior with the individual's data. The result is a posterior estimate of the parameter tailored to each individual (Fig. 2G). PEB has primarily been used in developing cancer screening algorithms that process longitudinal molecular biomarkers [4], [7], [9]. Specifically, the PEB screening algorithm involves computing the deviation of the biomarker level at each screening from a threshold that is adjusted to that individual. If the deviation is greater than a certain limit, the screening result would be positive. The individual-specific threshold is calculated based on the PEB estimator, a weighted average of the mean biomarker level in the control population, and the mean of previous biomarker levels for an individual. This value reflects both the general population trend and the individual's history of data. The prior, which is the mean biomarker level in the control population, is iteratively updated by the individual's sequential data. Using this algorithm on longitudinal protein levels to compute individual-specific screening thresholds, one study demonstrated that ovarian cancer can be detected earlier and at lower protein concentrations than a fixed threshold screening algorithm adjusted to the same specificity [4].

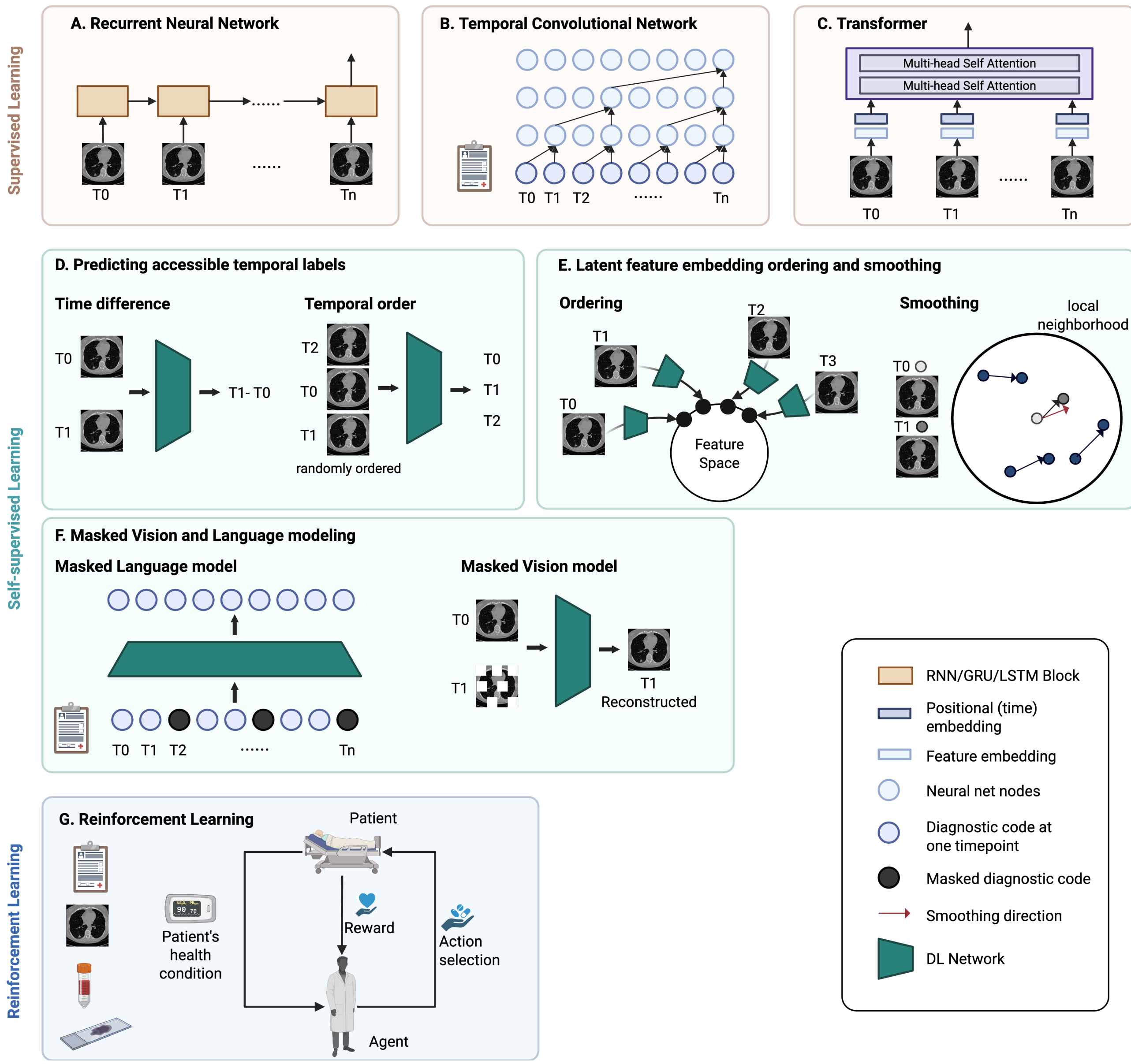

**Fig. 3. Deep learning models. Supervised learning** (A-C) uses labels to train the model. **(A)** In a recurrent neural network, the temporal dependencies in sequential data are modeled using a recurrent hidden state, which maintains the memory of previous states. **(B)** Temporal convolutional neural network uses causal convolutions, which convolve elements from the current and prior timepoints in the previous layer. Long-range dependencies can be achieved by the dilation mechanism in filters, which skips certain prior positions. **(C)** In a Transformer model, the multi-head self-attention mechanism helps the model learn different types of attention or dependencies between different parts of the longitudinal data. The order of the sequential data can be incorporated through positional encoding. **Self-supervised learning** (D-F) involves predicting either the data itself or easily accessible labels. This enables the acquisition of meaningful temporal information and serves as a robust feature extractor for downstream analysis, particularly in scenarios with limited data availability. **(D)** For example, self-supervised models can be trained to predict the time difference or the temporal order. **(E)** Adjacent time steps can be arranged closer together in the latent feature embedding, preserving the temporal relationship between data points. Also, the change trajectory can be smoothed by leveraging local neighborhood trajectories. **(F)** The models learn temporal dependencies by reconstructing corrupted clinical data or images from other timepoints. **(G) Reinforcement learning** algorithms learn the optimal actions to take in certain conditions and are iteratively optimized by reward functions. Created in BioRender. Hsu, W. (2025) https://BioRender.com/l80c624.

**Dynamic Bayesian network.** Dynamic Bayesian network (DBN) is a type of Bayesian network designed to model the temporal dependencies among variables. It depicts sequences of variables and their probabilistic interactions over time in a graphical structure. In this graphical representation, the nodes indicate the features, while the edges denote the probabilistic relationships between them (Fig. 2H). A DBN captures the state of each feature at a given timepoint, influenced by the features at the same timepoint and those at the preceding timepoint. This structure allows for the modeling of dynamic processes, where the state of a feature is a conditional probability of its current and immediate past states. DBNs have been widely used for modeling time series microarray data for gene regulatory network inference and cancer classification (cancer vs. non-cancer) [83], [84], [85]. They have also been investigated as cancer screening risk assessment models, using various longitudinal clinical and molecular features as nodes [86], [87]. Although DBNs inherently follow the Markov assumption, which states that the current state depends solely on the preceding state, they can have skip connections that allow the current state to be affected by an older timepoint [88].

**Hidden Markov model.** In the hidden Markov model (HMM), the temporal observations are dependent on hidden states, which are modeled using the Markov process. Expectation-maximization (EM) algorithm is implemented to estimate the parameters, such as transition probabilities and emission probabilities, to capture the temporal dynamics (Fig. 2I). HMMs are particularly well-suited for modeling the progression of diseases as hidden states [89], [90]. While symptoms are generally observable, the transitions between stages of disease progression are not always straightforward. Hence, HMMs can treat the observable symptoms as emissions and model their temporal patterns to investigate the underlying unobserved biological processes and disease stages. In addition, the Viterbi algorithm can be applied to estimate the most likely disease path for each patient, enabling personalized risk assessment and tailored treatment strategies. However, the characteristics of clinical time series data do not always satisfy the regular sampling and discrete format assumption of HMMs. A common approach to address this issue is to discretize the clinical time series data, but this can introduce errors and missing information. Continuous-time HMMs have been proposed as an alternative solution to allow transitions between hidden states and observations to occur continuously [91], [92], [93].

**Gaussian Process.** Gaussian process (GP) is a type of stochastic process that is ideal for modeling continuous time series data sampled irregularly (Fig. 2J). Its two main components are the mean function, which models the overall trend of the data, and the kernel function, which defines the correlation between timepoints. By providing a distribution of functions that could generate the data, GP can effectively capture non-linear relationships and make robust predictions while also providing uncertainty estimation. GP has been used to predict disease progression using clinical data, such as vital signs [94] and lab values [95]. In most studies, GP has been used to predict a single output from univariate or multivariate time series medical data. However, multi-task GP was also explored to learn the relationship between and within tasks simultaneously [96], [97], [98].

### *C. Deep learning*

DL involves training deep neural networks, which mimic how the human brain works. In the forward pass, neurons in one layer activate the neurons in the subsequent layer. The loss function calculates the difference between the predicted outcome and the actual value. The model learns by backpropagating the error and updating the weights to minimize the loss as much as possible. DL has found extensive application across various areas, especially in the medical domain, to facilitate decision-making [99]. With the rapid growth of medical data collection, neural networks have demonstrated high performance in handling unstructured, high-dimensional data. Additionally, DL has shown promise in processing time series data, excelling at capturing time dependencies without extensive preprocessing. In this section, we introduce three primary categories of DL approaches for analyzing temporal data: supervised, self-supervised, and reinforcement learning.

#### *1) Supervised learning*

**Recurrent neural network.** The recurrent neural network (RNN) is widely employed for modeling sequential data, leveraging recurrent hidden states to capture temporal dependencies and preserve the memory of preceding states (Fig. 3A). However, RNNs suffer from the gradient vanishing problem. Particularly, when the sequence is long, the gradients tend to diminish in the initial segments of the sequence. Hence, crucial information from the beginning may fail to propagate effectively throughout the network. Gated recurrent units (GRU) and long short-term memory (LSTM) have been proposed to address this short-term memory issue. Incorporating gating methods allows the model to retain and discard information over long sequences selectively.

RNN models have been extensively applied throughout medical imaging, molecular, and clinical data modalities. In medical imaging, deep and radiomics features are extracted from images at each timepoint [100], [101], [102], [103]. Molecular data, such as protein levels, are quantified from longitudinal serum samples [8], [50], [104]. Clinical data, including diagnosis codes, lab tests, and vital signs, are extracted and standardized [105], [106], [107]. In addition to vanilla RNN, bi-directional RNN models can replicate clinicians' approach to track patient history by observing disease progression in chronological order and then retracing it in reverse order to identify potential causes [108], [109].

**Temporal convolutional network.** An empirical evaluation has demonstrated that temporal convolutional network (TCN) models have considerable potential in sequence modeling and can outperform RNN models [110]. TCNs are an extension of CNNs, but the convolutions in TCNs are both causal and dilated (Fig. 3B). Causal convolutions prevent data leakage by constraining the dependence of present value solely on the past and current information. The receptive field, which refers to the number of past timepoints to be dependent on, can be adjusted flexibly. This hyperparameter can be determined based on domain expertise or by fine-tuning. On the other hand, dilated convolutions facilitate diverse receptive fields, which capture time dependencies across different time ranges. TCNs are utilized for early disease diagnosis and disease progression modeling across both clinical and imaging modalities [66], [111], [112], [113]. Within TCNs, hierarchical attention mechanisms are leveraged to acquire the optimal representations at various stages, enhancing the interpretability of the model [112], [113].

TABLE I
ADVANTAGES AND DISADVANTAGES OF LONGITUDINAL MODELING TECHNIQUES.

| Method[a] | Pros | Cons |
|---|---|---|
| | **Feature Engineering** | |
| **CAT** | • Simple method to feed the downstream model all datapoints across time. | • Increases dimensionality of data, leading to overfitting.<br>• Does not provide the model with information on the duration or change between timepoints. |
| **PD** | • High interpretability – intuitive what kind of temporal information is being utilized. | • Impedes the discovery of new and potentially important features that are not perceivable by humans since features are pre-defined. |
| **TRANS** | • Extracted features can characterize complex longitudinal patterns in the data trajectory, compared to handcrafted features. | • Transformed features are not intuitive.<br>• Requires many timepoints to accurately interpolate trajectories and characterize underlying trends. |
| **SIM** | • High interpretability – simple assumption that similarly shaped trajectories share the same outcome.<br>• Robust to varying alignment and length of temporal data. | • Requires many timepoints to accurately interpolate trajectories and identify detailed shapes to distinguish classes. |
| | **Statistical Model** | |
| **MEM** | • Inherently robust to missing and irregularly spaced data, as it leverages all available data to estimate its parameters based on the maximum likelihood of the observed data. | • Requires a priori selection of parametric forms to model random effects and relationships between variables. |
| **PEB** | • Does not involve many assumptions compared to more complex statistical models.<br>• Does not require model estimation to characterize biomarker trajectory, making it suitable for data with limited timepoint. | • Not appropriate to use on temporal data with varying time intervals, as it is unaware of the rate of change in biomarker level over time. |
| **DBN** | • High interpretability – graphical structure that visually represents feature connections and quantifies their relationship with conditional probabilities. | • Assumes a fixed time interval between observations.<br>• Lacks flexibility in handling additional timepoints since it is parametrized based on a specific number of timepoints.<br>• Highly sensitive to the quality of domain knowledge used to form relationships between variables. |
| **HMM** | • Suitable for modeling disease progression in an unsupervised manner, without labeled data.<br>• Uncovers hidden disease path on the patient level. | Requires strong assumptions between prior and current states.<br>• Fails to capture long-range dependencies due to its Markov assumption. |
| **GP** | • Provides an uncertainty estimation, which evaluates the reliability or confidence of the model in the prediction.<br>• Flexible in handling inconsistently sampled data. | • Requires careful selection and tuning of the kernel. |
| | **Deep Learning** | |
| **RNN** | • Relatively memory efficient compared to the Transformer.<br>• Performs well in smaller datasets as they have fewer parameters than the Transformer. | • Assumes a fixed time interval between observations.<br>• Data processed one timepoint at a time, resulting in a long training period.<br>• Gradient explosion or vanishing during backpropagation. |
| **TCN** | • Generally faster to train and less likely to have vanished gradients, compared to RNNs. | • Extensive hyperparameter tuning on receptive field and dilation necessary to achieve optimal results. |
| **TF** | • Effectively handles long-range dependencies.<br>• Learns rich dependencies across timepoints through the attention mechanism.<br>• Training time is faster due to parallelization. | • Requires a large amount of training data due to its high number of parameters.<br>• Computationally expensive due to the reliance on the attention mechanism |
| **SSL** | • Does not rely on labels or annotations, which are limited in medical data. | • Requires large datasets for effective training.<br>• Requires careful design of pretext tasks to learn relevant representations. |
| **RL** | • Addresses a unique task – provides guidance on treatment and intervention planning in real time. | • Unclear environment – disease progression is a complex process with unknown biological pathways and interactions.<br>• Medical data do not contain all possible state-action pairs for training.<br>• Requires careful consideration in designing reward functions. |

[a]CAT = Concatenation, PD = Pre-defined Features, TRANS = Transformation, SIM = Similarity Calculation, MEM = Mixed Effect Model, PEB = Parametric Empirical Bayes, DBN = Dynamic Bayesian Network, HMM = Hidden Markov Model, GP = Gaussian Process, RNN = Recurrent Neural Network, TF = Transformer, SSL = Self-supervised Learning, RL = Reinforcement Learning.

**Transformer.** The Transformer, first introduced in 2017 for natural language processing tasks, utilizes attention mechanisms to learn sequence dependencies [114]. The core components in the Transformer that are particularly suitable for modeling sequences include positional encoding and multi-head self-attention (Fig. 3C). The positional encoding provides information on the order of the sequential data. It can also be adapted for time series data to represent temporal information. Secondly, the self-attention mechanism allows the model to capture dependencies between all elements in the sequence, regardless of their positions, preventing prior information from being lost in the training process. Multiple self-attention modules can be stacked to form multi-head attention, each capturing different types of relationships within the sequence. In longitudinal medical imaging research, features are extracted from medical images at each timepoint and input into the Transformer model [102], [115], [116], [117], [118]. Time can be encoded using the actual time value or the intervals between images.

*2) Self-supervised learning*

Utilizing supervised models with medical data poses significant challenges, primarily due to the difficulty of obtaining accurate labels or annotations for the data. In most supervised learning approaches, a lot of unlabeled medical data is being underused. Self-supervised learning offers a promising approach to enable the model to extract valuable temporal representations from the data without relying on explicit labels or annotations. However, large, high-quality datasets are often required for self-supervised methods to perform effectively. There are several proxy tasks that can be used to train self-supervised models to learn temporal information.

**Predict accessible temporal labels.** Self-supervised models can be trained to predict the temporal order or the time interval between each element in the data sequence [119], [120] (Fig. 3D). In this way, models learn how anomaly patterns evolve, which enhances the effectiveness of supervised learning performed concurrently or downstream.

**Latent feature embedding ordering and smoothing.** Contrastive learning learns embeddings by pulling adjacent time steps of the same time series together. Therefore, the latent embedding space preserves the temporal relationship between data points [121]. In another approach, morphological change trajectories are smoothed using local neighborhood embedding. The resulting informative latent space representation effectively differentiates diagnostic groups even in the absence of labeled training data [122] (Fig. 3E).

**Masked Vision and Language Modeling.** Bidirectional Encoder Representations from Transformers (BERT) is a Transformer-based pre-training model that uses masked language modeling (MLM). BERT has been adapted to structured EHR data and has shown substantial improvements in predicting disease outcomes in downstream tasks [123], [124], [125], [126]. In MLM, certain elements in the input sequence are randomly masked out, and the model is trained to predict the original value of the masked element. The inclusion of temporal information is achieved through time embedding and visit segment embedding. Similarly, in medical imaging studies, image reconstruction serves as a crucial technique in self-supervised pretraining, where models are trained to reconstruct corrupted input images. In longitudinal studies, the model reconstructs the original image using not only the local contextual information but also medical images from other timepoints [127] (Fig. 3F).

*3) Reinforcement learning*

While most methods presented thus far perform diagnostic or prognostic tasks, reinforcement learning aims to provide guidance on what treatment and intervention could lead to better patient outcomes at each time point. In reinforcement learning, an agent learns to iteratively take actions (e.g., treatments) that interact with the dynamic environment (e.g., patient status) and maximize the rewards (e.g., patient outcome) (Fig. 3H). Several recent studies have leveraged sequential medical imaging and clinical data to optimize treatment and screening policies using this approach [128], [129], [130], [131].

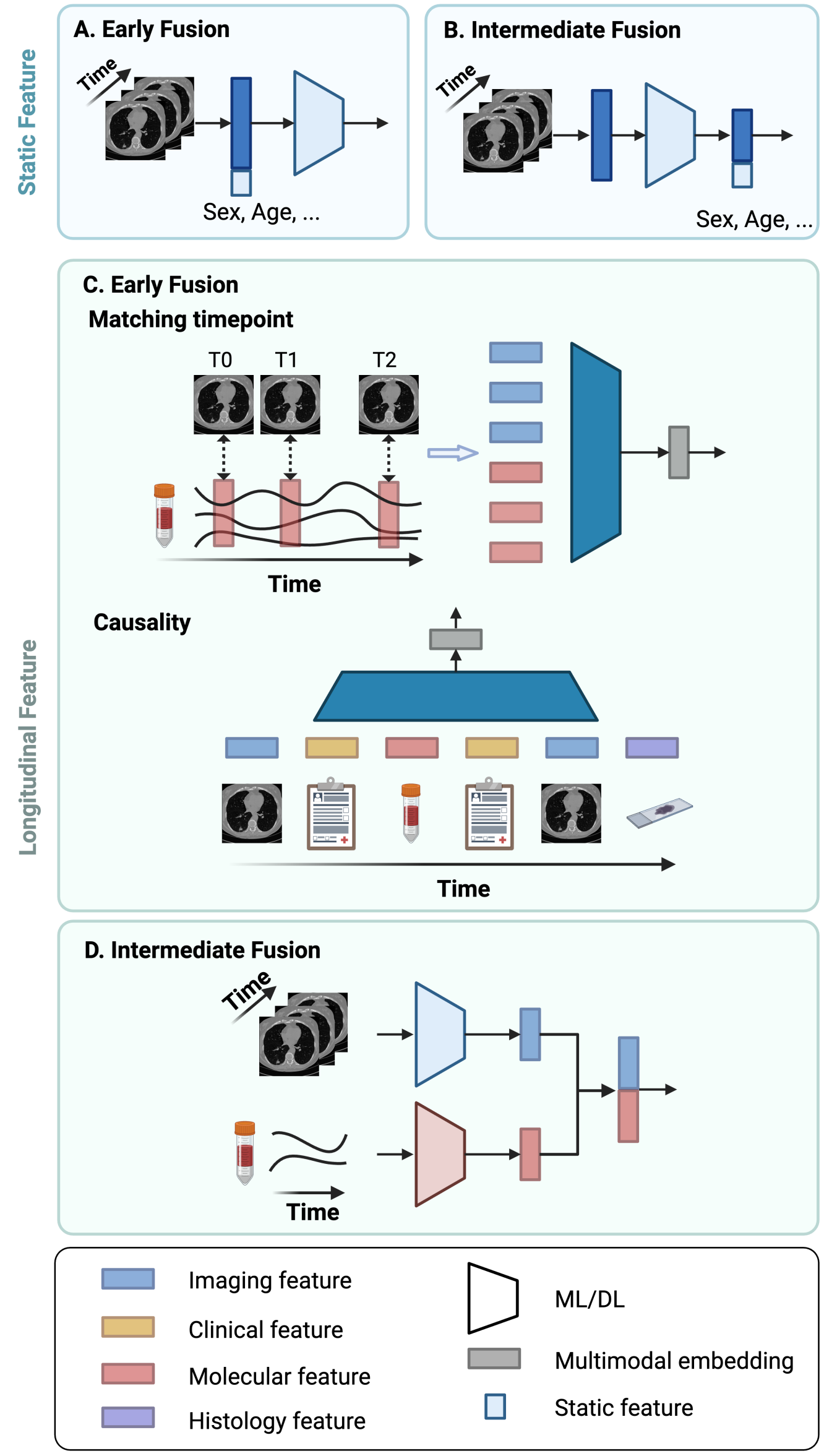

**Fig. 4. Multimodal longitudinal modeling.** Longitudinal data are typically fused with static features (A-B), such as demographics, or another longitudinal modality (C-E). When fusing with longitudinal features, the static feature can be concatenated either **(A)** before or **(B)** after passing onto the machine learning (ML)/ deep learning (DL) model. **(D)** If all the modalities contain longitudinal features, each modality can be modeled separately, and the feature embeddings can be fused at the end. **(C)** There are two ways to perform early fusion of longitudinal data. First, features at the matching timepoint can be selected and passed onto DL models for temporal analysis. Second, all features from all modalities can be arranged in the order of causality before feeding into DL models. Created in BioRender. Hsu, W. (2025) https://BioRender.com/t46h753.

## IV. Multimodal Modeling

Historically, cancer research has focused on analyzing data from a single modality. However, there has been a surge of interest in developing multimodal fusion techniques for heterogeneous medical data in recent years. In oncology, multimodal fusion can use the synergistic and complementary contextual information from diverse modalities to provide a comprehensive understanding of cancer.

When applied longitudinally, these fusion methods can uncover temporal dependencies across different modalities, potentially enhancing the predictive power of diagnostic and prognostic models. Detailed descriptions of multimodal modeling with various fusion methods (early fusion, intermediate fusion, late fusion, and attention-guided fusion) have been presented in numerous review papers [17], [18], [132]. However, longitudinal modeling of multimodal data remains relatively under-investigated in current research.

**Integration with features at a single timepoint.** Longitudinal features are often integrated with static features, such as demographics (e.g., gender, age, race). At the beginning of a clinical study, surveys can be administered to gather personal and family health histories, such as previous diagnoses and cancer occurrences among relatives. These static features can serve as additional risk factors, and their integration with longitudinal data allows for a more comprehensive assessment of an individual's cancer risk. In ML, static features are typically concatenated with longitudinal features in early fusion [44] or integrated with deep features using intermediate fusion [133] (Fig. 4A-B). Here, concatenation occurs across different modalities, in contrast to longitudinal concatenation, where features are combined across timepoints.

**Integration with longitudinal features.** When integrating longitudinal features from different modalities, modalities can be collected simultaneously or in close succession. For instance, in MRI, multiple series of images, such as T1-weighted, T2-weighted, diffusion-weighted imaging (DWI), and dynamic contrast-enhanced (DCE), are collected only within a few minutes of each other. Typically, studies utilizing such data concatenate the features across matching timepoints for multimodal modeling [102], [134]. However, in the medical domain, it is not always the norm for modalities to be collected simultaneously. More often, different modalities of medical data are collected at varying sampling rates and time scales, requiring sophisticated techniques for harmonization and temporal alignment. Additionally, each modality can present unique issues, such as missing data. To address the issue, one study extracts clinical features only on dates that correspond to those of imaging features [118] (Fig. 4C). These selected features are then input into a Transformer model, with positional embeddings to denote the order of the feature and segment embeddings to denote feature modalities. However, this approach can result in underutilizing clinical features, potentially relevant risk factors.

In other studies, the entire sequence of medical data from multiple modalities is considered. For example, the modalities are treated independently and separately modeled, after which the resulting features are aggregated by concatenation [135], [136] (Fig. 4D). However, this approach may overlook the interactions between different modalities at various timepoints. In another study, the causal relationship between and within modalities is considered [137] (Fig. 4C). For instance, vital signs, lab values, and imaging can collectively contribute to patients' diagnoses, subsequently influencing the selection of procedures and medications for their treatment. Therefore, features from each modality are organized and aligned sequentially before being fed into an RNN model. The issue of varying feature lengths is addressed by applying zero-padding.

## V. Applications of Longitudinal and Multimodal Modeling in Precision Oncology

Precision oncology involves personalized methods for diagnoses, treatment, and monitoring that are specifically tailored to the unique characteristics of each patient's cancer. These characteristics are typically reflected in various types of data gathered throughout the patient's journey (Fig. 1A). In this section, we highlight the existing literature and potential of using longitudinal and multimodal modeling methods to advance precision oncology, supporting diagnosis and prognostic tasks such as treatment response, recurrence, and survival prediction (Fig. 1B).

### *A. Diagnosis*

Early detection through screening and accurate diagnosis of cancer enables timely intervention and improved outcomes [138]. Longitudinal modeling enhances these tasks by capturing subtle temporal changes in disease-related features. A range of longitudinal modeling approaches has been explored to support these applications.

Feature engineering techniques such as concatenation [25], [43], [45], [47] and extraction of predefined features [3], [5], [24], [26], [51] have been employed. For example, time series laboratory test results from EHR are concatenated and processed with neural networks for early detection of pancreatic cancer [45], [47]. As for statistical models, MEM [6], [77], [78], [79], PEB [4], [7], [9], DBN [84], [86], [87], [88], and HMM [93] have been successfully applied. Prominent examples include the use of a Bayesian hierarchical mixture change-point model to analyze serial protein levels for improving screening sensitivity [6], [77], [78], and PEB approaches to derive personalized biomarker thresholds for screening [4], [7], [9]. More recently, DL approaches, including RNN [8], [100], [101], [103], [104], Transformer [117], and reinforcement learning [129], [130] have demonstrated promise. For instance, LSTM is used to process longitudinal CT for enhanced lung cancer early detection [101], [103]. Notably, reinforcement learning can be used for optimizing breast cancer screening intervals based on AI-derived risk scores from mammograms [129].

Regarding multimodal modeling of temporal data, early fusion of longitudinal clinical and medical imaging data in matching timepoints has exhibited enhanced efficacy in cancer diagnosis [118]. The integration of other combinations of longitudinal data modalities, such as imaging and molecular biomarkers, still remains underexplored.

### *B. Prognosis*

Prognostic modeling is central to precision oncology, enabling patient stratification, treatment personalization, and dynamic risk assessment over the disease course. Longitudinal data are valuable for prognosis, as they capture how disease features evolve in response to therapy or progression. Various modeling approaches aid in leveraging longitudinal imaging and molecular biomarkers collected throughout patient care to predict tumor progression [89], treatment response [11], [39], [40], [44], [49], [50], [52], [53], [54], risk of recurrence [55], [74], [76], [80], and survival [49], [57], [58], [59], [60], [139].

Extraction of predefined features is particularly useful for these prognostic tasks [11], [40], [49], [50], [52], [53], [54], [55], [56]. For example, delta-radiomics, the difference in radiomics features between two imaging timepoints, is used to predict response and survival in lung cancer patients treated with immune checkpoint inhibitors [49]. Another study uses postoperative serial circulating tumor DNA to predict recurrence in patients undergoing resection of colorectal cancer liver metastases [55]. Moreover, FPCA is actively used for prognosis, especially in survival prediction tasks involving longitudinal molecular biomarkers [57], [58], [59], [60]. Statistical models, such as MEM [74], [76], [80] and HMM [89], have been widely studied. A common application involves predicting prostate cancer recurrence using MEMs jointly modeled with risk models to process repeated measurements of prostate-specific antigen [74], [76], [80]. While less common, DL models like RNN have also been explored for prognostic tasks [50], [140].

Prognosis is a complex task that encompasses various facets of cancer, including genomic profiles, cancer subtypes, and tumor morphology. Therefore, multimodal fusion techniques are often employed to improve treatment response prediction [20], [21], [22]. However, most existing fusion models overlook the importance of longitudinal changes in individual modality and the temporal relationships across modalities, both of which are essential for enhancing accuracy and robustness in prognostic tasks. A recent study used early fusion to integrate longitudinal MRI with single-timepoint clinical and molecular data to improve chemotherapy response prediction in breast cancer patients [44]. Such longitudinal multimodal approaches should be investigated more extensively in future research.

## VI. Challenges and Future Directions

Research in the nascent area of developing longitudinal multimodal biomarkers has thus far demonstrated potential in advancing precision medicine. Nevertheless, further efforts are needed to refine and promote longitudinal multimodal analysis in cancer research. Many considerations and challenges exist throughout the entire pipeline of longitudinal multimodal cancer research. In this section, we highlight these challenges, discuss existing and potential approaches to address them, and suggest future directions to advance the field.

### *A. Longitudinal data acquisition and sharing*

Data sharing involves making data accessible to others to promote transparency and enable further analysis. This practice can significantly accelerate the pace of biomarker discovery and predictive model development while enhancing the reproducibility and generalizability of studies. Hence, considerable efforts have been made to promote data sharing, aiming to foster a more collaborative and open scientific community. Public data repositories serve as platforms for facilitating data sharing, allowing researchers access to diverse datasets. Notable examples include The Cancer Genome Atlas [141] and The Cancer Imaging Archive [142], both of which have been pivotal in advancing research in the genomic and imaging domains, respectively. However, despite the promotion of data-sharing practices and the availability of public repositories, longitudinal datasets in oncology remain limited.

Several hurdles may be involved in acquiring and sharing longitudinal data in oncology. One major challenge is the complexity of collecting longitudinal data, which involves multiple follow-ups with patients over extended periods. This can be resource-intensive and time-consuming. Additionally, patient dropout and loss to follow-up can prevent the collection of large numbers of complete samples. Some ways to address this are to enhance patient engagement and communication, offer incentives, or provide flexible scheduling options for follow-up [143]. Although not unique to longitudinal data, privacy concerns for patient health information pose substantial barriers to data sharing in the medical domain. De-identification of patient data, consent management, access control, and data encryption are critical measures to ensure secure data sharing. Another solution could be to utilize federated learning, which allows data to remain at their host institutions while being available for analysis within secure networks [144], [145]. Continued investment in the infrastructure of public data repositories will also be essential to promote the sharing of longitudinal datasets.

Notably, in medical fields other than oncology, longitudinal data collection has been ongoing for many years. Studies such as Alzheimer's Disease Neuroimaging Initiative [146] and Parkinson's Progression Markers Initiative [147] recruit both healthy individuals and those with various disease conditions, tracking them over long periods. They collect a wide range of data, including imaging, clinical data (e.g., vitals, diagnosis, and medication), blood tests, genetic information, and surveys. Thousands of papers have been published using these open-source datasets to identify biomarkers for diagnosis and evaluate disease progression. Such large-scale longitudinal biomedical data databases would not be possible without support from the government, companies, and researchers who contribute their expertise and resources. In oncology, a similar approach should be benchmarked so that comprehensive longitudinal data can be readily accessed for better biomarker discovery and improved patient outcomes.

### *B. Intra- and inter-subject variability*

Each data modality confronts unique challenges that induce confounding factors, hindering the validity and robustness of models. Variability in clinical data may arise from differences in clinical procedures across various providers. For molecular data, variations in the storage and processing of biospecimens, methods of measuring analytes, and bioinformatics algorithms for generating features contribute to data heterogeneity. Likewise, in imaging, variability stems from differences in equipment, image acquisition parameters, and reconstruction settings. These systematic variabilities are commonly referred to as batch effects and can be more prominent with longitudinal data, which involves multiple batches of data collection. Hence, when developing longitudinal models, preprocessing to remove or minimize batch effects is critical in addressing these issues.

Longitudinal studies require careful consideration of intra-subject variability across multiple timepoints, as well as inter-subject differences. For all modalities, standardization or normalization of data adjusts values to a standard scale and range, which helps alleviate variations across different collection points. This preprocessing technique is also critical when dealing with multiple features with varying scales or multimodal features, especially when using certain ML or DL models. Batch correction methods like ComBat can be considered to address batch effects in molecular data and imaging data [43]. In addition, DL models, such as generative-adversarial networks (GANs), variational autoencoders (VAEs), and diffusion models, have found extensive use in medical imaging to enhance image consistency through harmonization [148]. Another preprocessing technique that is essential for sequential imaging data is registration. Registration aligns images from different time points to a single coordinate system, ensuring that comparisons and measurements are made on anatomically corresponding regions.

### *C. Model training under limited data and label constraints*

The limited availability of longitudinal datasets in oncology leads to issues in model training, especially for DL models that require a large amount of data. A potential solution is transfer learning, which takes a model pre-trained on a large dataset and fine-tunes it on a smaller task-specific dataset. Emerging foundation models trained on various data types [149] can serve as effective feature extractors, potentially leading to improved downstream performance in longitudinal analysis.

Cross-sectional data, which is more abundant, can also be used to model underlying dynamic processes. Pseudotime analysis has been extensively used in single-cell RNA sequencing to study cell differentiation and development, assuming the cells sampled can be arranged along a biological trajectory. Multiple trajectory inference techniques have been proposed to assign pseudotime by analyzing the similarities in the expression patterns [150], [151]. Likewise, they can be applied in other medical domains, such as medical imaging and computational pathology, where the morphology observed in different cases can represent the continuum of states along disease development.

Acquiring labels, especially in oncology, presents significant challenges as annotations typically require skilled clinicians. This process becomes even more demanding and time-consuming for pixel-level annotations on imaging data. Also, the manual annotation process inherently introduces the potential for noise and human bias. As mentioned earlier in Section III, self-supervised learning offers a promising solution. This approach allows models to leverage a larger pool of unlabeled temporal data for pretraining and learn temporal features by predicting the data itself or utilizing more readily accessible labels, like time intervals.

Furthermore, large language models (LLMs), which are trained on extensive and diverse datasets, can capture nuanced dependencies and contextual meanings effectively. Although LLMs are typically applied to text data, research has shown that LLMs also possess some capability to interpret time series data [152]. Specifically, with domain-specific prompt engineering and prompt tuning, LLMs can serve as effective few-shot health learners, even with limited data [153].

### *D. Inconsistent time interval*

Outside clinical trials, time series data are usually collected at irregular intervals. In cancer screening, for example, patients at higher risk or displaying concerning screening results typically undergo more frequent follow-up appointments. Additionally, longitudinal data exhibit numerous missing values for reasons such as patient dropout or missed appointments. This diversity poses unique challenges in analyzing and modeling temporal data derived from clinical studies. While some statistical modeling approaches, such as MEM and GP, inherently handle and impute irregularly spaced time series data, most longitudinal modeling methods require the input data to be structured vectors with uniform lengths and no missing values.

To address this, many studies have opted to design inclusion criteria to select a predetermined time frame or a fixed number of timepoints to be fed into the model. However, such a strategy can result in underutilization of available data and biased prediction due to the excluded data. Numerous preprocessing techniques have been proposed to standardize time series data, ensuring uniform time intervals between sampled time points. Imputation is the most prevalent method for handling missing values. Various approaches exist, including simple padding with zeros and forward or backward filling [8], [45], [135]. Some works also replace missing values with either the median or the mean value of the observed data [48]. More sophisticated imputation methods (e.g., KNN, multivariate imputation by chained equations) take the underlying structure and relationship into account to predict the missing values. However, these methods can be biased if the data are not missing completely at random. Other approaches treat the missingness itself as an indicator or a variable in the analysis [8], [88], [154].

Another approach to address the irregularity of time intervals in healthcare data is to incorporate time information into the model explicitly. Both the time interval and time to the latest observation can be integrated into temporal models. One method involves concatenating the time differences between records to the feature embedding [47]. In RNN, the short-term memory is adjusted based on the elapsed time between records, where longer time gaps lead to a reduction in short-term memory [155], [156]. While such a method focuses on the local time intervals, the temporal emphasis model (TEM) serves as a global multiplicative function applicable to both the RNN and Transformer self-attention module [101], [117]. It computes the time relative to the latest observation and scales the importance of each timepoint accordingly. TEM effectively prioritizes the most relevant and recent information for prediction, minimizing the impact of outdated data on the model's performance. In addition, the neural ordinary differential equation offers a framework to parameterize the continuous dynamics of hidden states within neural networks. Since time is treated as a continuous variable in ODE, it can handle irregularly sampled temporal data and is especially suitable for modeling disease progression [139], [157]. Moreover, the GP adapter allows end-to-end training with backpropagation [158]. The parameters in the GP adapter can be trained along with the DL model (e.g., RNN and TCN) to provide precise data imputation tailored to the specific task and guarantee a uniform format of input for DL models in disease prediction [66], [107].

Despite the various approaches to tackle the problem of data collection irregularity, the best approach remains to be determined. It will be beneficial to conduct comparisons based on different data types to determine the most effective method for specific scenarios. Additionally, the current methods may lack practicality in clinical settings. For instance, although time-aware RNNs solve the issue of irregular time intervals, they still require the same number of timepoints when training in batches. Imputing data for new patients with limited data to match the length of data for existing patients is illogical and could introduce bias in the imputation process. Moreover, the practice of adding zeros raises concerns regarding how the model interprets these zeros within the data sequence and may result in suboptimal performance. The complexity of the issue escalates when dealing with missing medical images, which would require training additional models for image generation. Hence, there is a need for more adaptable and flexible approaches that cater to patient data with a varying number of timepoints.

### *E. Temporal ordering and alignment in multimodal fusion*

Given that an individual is enrolled in a screening program, the patient's medical history and demographic information are collected, followed by blood tests and imaging scans. These initial screenings serve as a preventive measure to detect any potential abnormalities related to the early signs of cancer. Once a concerning finding has been identified, the patient may undergo more diagnostic imaging and biopsies to obtain a final diagnosis. During the treatment of cancer, imaging scans and laboratory tests are performed to further monitor the cancer progression and the patient's response to treatment. While some data can be collected at the same time, most of them are acquired at various timepoints. In some recent multimodal studies, there is a concerning trend where temporal information is completely disregarded during the process of aggregating multimodal data. This could lead to significant gaps in understanding the dynamic trend of cancer progression. On the other hand, some studies tend to synchronize the exact number of timepoints between different modalities. While such an approach solves the issue of time alignment, it may result in the underutilization of available medical data and lead to biased predictions. Hence, there is a growing need for research to develop practical approaches to integrate medical data in a manner that allows for the incorporation of temporal information. In addition, the intrinsic temporal information present in multimodal data not only provides insights into disease progression but also suggests potential causal relationships. For instance, genetic mutations can lead to various morphologic changes in the imaging features, such as tumor growth patterns and the tumor microenvironment. These distinct modalities, captured at different times, collectively contribute to predicting treatment responses and survival outcomes. Simply combining all modalities can lead to highly biased estimation [159]. Therefore, integrating data from different modalities, while also accounting for causal relationships across time, necessitates more sophisticated methodologies.

### *F. Determining the importance of time period and modality*

Another challenge in multimodal fusion involves the weighting strategies used to integrate different types of medical data. Some studies have proposed methods to understand the interaction between various modalities by quantifying the redundancy, uniqueness, and synergy levels, offering insights into optimal fusion strategies needed for the task [160]. Other methods enable the model to learn the weights of each modality, but this approach can result in model decisions being dominated by a single modality. For example, genomics features often play a more significant role in predicting cancer patients' survival than radiology features, as particular mutations are highly correlated with treatment outcome and prognosis. The model would learn the relationship within genomics features more quickly, thereby neglecting the potential for weaker modalities to reach their full training capacity. Several methodologies, including ensemble, attribution regularization, and modification of the learning process, have been proposed to tackle the problem of modality imbalance and make full use of the data [161], [162]. Similarly, in temporal analyses, determining which time period requires greater emphasis is uncertain. Typically, recent visits and scans are often given more weight because they reflect the most current state of the patient's health. However, this approach may not always be appropriate. The patient's baseline health conditions and comorbidities from earlier times can have a significant impact on the treatment decisions. Therefore, domain knowledge and careful planning are essential to tailor to each specific cancer type and prediction task.

### *G. Longitudinal model selection*

In Section III, we detail different longitudinal modeling approaches, yet the question remains: which method is superior? Determining the most appropriate method for various longitudinal datasets and clinical contexts can be challenging. To assist researchers in this process, Table I summarizes the strengths and limitations of each method.

A general pattern can be observed in how previous studies have applied these methodologies. Recently, DL models have gained popularity in temporal data analysis due to their ability to uncover complex relationships in high-dimensional and long-sequence datasets. Accordingly, clinical data such as vital signs or lab tests are typically modeled using DL approaches, since they are gathered frequently for extended periods. Additionally, DL models excel at capturing spatial and textural relationships in images, making them ideal for handling longitudinal medical imaging data. However, these models often require greater computational resources, including high memory and Graphics Processing Unit (GPU) capacity, and tend to have longer training times.

In contrast, feature engineering and statistical models provide more robust predictions when working with smaller datasets and fewer features. These models generally tend to be more interpretable since they are either inherently explainable or can leverage post-hoc explanation algorithms, as discussed in Section H. Feature engineering and statistical techniques are typically applied to longitudinal handcrafted features extracted from imaging data or molecular biomarkers, which usually involve only one or a few variables. Many statistical models require certain assumptions to be fulfilled, which can be advantageous or disadvantageous depending on the understanding of the biological implications and the relevant domain knowledge.

Despite the observed trends, there is still a lack of comprehensive benchmarks. Most studies compare the proposed methods only to state-of-the-art models within their respective domains, and very few compare results across feature engineering, statistical methods, and DL models simultaneously. Developing such benchmarks would require collaboration among experts from different domains, offering deeper insights into the effectiveness of these models and promoting innovation in longitudinal modeling across diverse clinical contexts.

### *H. Explainability*

Feature engineering approaches to longitudinal modeling have relatively high interpretability because many of the features are handcrafted and explicitly encode time information. The importance of each feature can be obtained by the learned coefficients or using explainability methods, such as SHapley Additive exPlanations [163]. Statistical models are also highly interpretable, as they usually deal with low-dimensional features and have clearly defined parameters and relationships. The issue with interpretability primarily arises with DL models. In DL models trained on medical imaging data, gradient-weighted class activation mapping [164] or attention scores from the attention mechanism are utilized to indicate the importance of different parts of the input data to the model's decision-making process. However, the attention maps do not inherently suggest what characteristics high-attention regions represent and whether they positively or negatively contribute to the prediction.

Explaining models that leverage longitudinal medical imaging presents an additional challenge. While specific studies have demonstrated improvement in diagnostic or prognostic tasks by integrating prior imaging scans, the exact reason behind this improvement remains unclear. It is uncertain whether this enhancement is solely due to the increased model complexity or if the model effectively captures changes in the ROI. Overlaying heatmaps or attributions generated by explainability methods or attention modules does not provide additional clarity or insight [45], [100], [116]. In addition, multimodal features often have high correlations and complex interactions, especially in the temporal setting. Traditional explanation techniques struggle to handle interactions between features, leading to a flawed understanding of feature importance. While most studies analyze the importance of each modality independently, they often overlook the causal interactions in multimodal fusion. It is worth exploring how alterations in one modality might induce changes in another, thereby collectively influencing the outcome.

### *I. Evaluation of longitudinal and multimodal models*

The performance of longitudinal multimodal models is evaluated based on metrics that are commonly used for specific tasks. For instance, the area under the receiver operating characteristic (AUROC) curve is often used in classification tasks, whereas the hazard ratio is used for risk analysis tasks. However, since longitudinal multimodal models have significantly greater data requirements and are more complex than models that use single timepoints and modalities, additional evaluation must be performed to justify the longitudinal and multimodal components.

One way to conduct this evaluation is to compare the performance of longitudinal multimodal models against single-timepoint or single-modality baseline models [3], [4], [5], [6], [7], [9]. This would clearly demonstrate the value of incorporating multiple timepoints or modalities. Another point of interest may be to evaluate the impact of individual timepoints or individual modalities on model performance. Ablation studies can be useful for this purpose. Systematically varying the number of timepoints can help quantify how additional timepoints contribute to the overall performance of the model [8]. For example, showing that adding subsequent timepoints improves the AUROC of a classification task can substantiate the benefits of each individual timepoint. For models that have time-encoding components, the model performance with and without the component can be compared [135]. Similar analysis can be performed to validate the inclusion of data from multiple modalities and demonstrate the contribution of each. An important consideration is that the value of increased performance should be significant enough to outweigh the difficulties associated with obtaining longitudinal and multimodal data. A standardized framework that can help quantify this significance and derive conclusions about the optimal number of timepoints or modalities is needed.

## VII. Conclusion

Longitudinal and multimodal analysis is crucial for advancing precision oncology through a comprehensive assessment of cancer complexity and heterogeneity. In this review, we introduced a broad range of methods for modeling longitudinal and multimodal data and conveyed their promise for enhancing cancer research. However, significant challenges still remain, including the need for improved data collection and sharing, integration techniques for longitudinal data from different sources, model explainability for clinical decision support, and standardization of the evaluation process. Addressing these challenges will be essential to achieving early cancer detection and developing personalized treatment strategies, ultimately leading to better outcomes for cancer patients.

SUPPLEMENTAL TABLE I

A SUMMARY OF PUBLICLY AVAILABLE LONGITUDINAL DATASETS IN CANCER.

| Datasets[a] | Cancer Type | Study Type | Modalities | | | |
|---|---|---|---|---|---|---|
| | | | Clinical | Molecular[b] | Imaging[c] | Histology |
| NLST [1] | Lung cancer | Clinical trial: Screening | Demographics, medical/smoking/family history, diagnosis, treatment, survival | Biospecimens: blood, sputum, urine, tissue | Low Dose CT, Chest X-ray | Whole Slide Image |
| ACRIN 6668 Multi-center Clinical Trial [2] | Non-small cell lung cancer | Clinical trial: Response to chemoradiotherapy | Demographics, lab test, diagnosis, treatment, survival | - | FDG-PET, CT | - |
| PLCO Cancer Screening Trial [3] | Prostate, lung, colorectal, and ovarian cancer | Clinical trial: Screening | Demographics, medical/smoking/family history, medical history, surveys, lab test, diagnosis, treatment, survival | Biospecimens: blood, tissue<br>Biospecimen results: CA125 level, GWAS, WGS, WES, tumor sequencing, EWAS, serum metabolomics, oral microbiome | Chest X-ray | Whole Slide Image |
| UKLWC [4] | Ovarian cancer | Clinical trial: Screening | Demographics, medical/smoking/family history, diagnosis, treatment, survival | Biospecimens: blood<br>Biospecimen results: CA125 level | US | - |
| IMpower150 [5] | Non-small cell lung cancer | Clinical trial: Response to drugs | Demographics, smoking history, lab test, radiology report, survival | ctDNA data | - | - |
| The Cancer Moonshot Biobank [6] | Cancer: colon, lung, prostate, breast, ovarian, melanoma, gastroesophageal, myeloma, acute myeloid leukemia | Biobank | Demographics, lab test, diagnosis, treatment, survival | Biospecimens: blood, tissue<br>Biospecimen results: SNP, CNV | CT, MRI, PET, US, Angiography, X-ray | Whole Slide Image |
| CSAW-CC [7] | Breast cancer | Clinical trial: Screening | Demographics, medical history, diagnosis, treatment, survival | - | X-ray Mammography | - |
| I-SPY TRIAL [8] | Breast cancer | Clinical trial: Response to drugs | Demographics, immunohistochemistry report, surveys | Biospecimens: blood, tissue | MRI | Core-needle biopsies |
| EA1141 [9] | Breast cancer | Clinical trial: Screening | Demographics, medical/family history | Mutation | MRI, DBT | - |
| QIN Breast [10] | Breast cancer | Clinical trial: Response to neoadjuvant chemotherapy | Treatment response | - | PET/CT, DCE-MRI | - |
| ReMIND [11] | Brain tumor | Surgical resection | Demographics, medical history, diagnosis | Mutation | MRI, US | - |
| PEDSnet [12] | Brain tumor, leukemia, lymphoma | Biobank | Demographics, lab test, treatment, medication, diagnosis, outpatient/inpatient/Emergency visits | - | - | - |
| UK Biobank [13] | All cancer types | Biobank | Demographics, lab tests, surveys, activity monitoring, primary care data, inpatient data | Biospecimens: blood, urine, saliva<br>Biospecimen results: WGS, WES, Genotype data | MRI | - |
| The All of Us [14] | All cancer types | Biobank | Demographics, vital sign, lab test, medical procedure, diagnosis, surveys, digital health | Biospecimens: blood, saliva, urine<br>Biospecimen results: WGS | - | - |
| MIMIC [15] | All cancer types | Intensive Care Unit | Demographics, vital sign, lab test, medical procedure, medication, clinical notes | - | Chest X-ray | - |
| eICU [16] | All cancer types | Intensive Care Unit | Demographics, vital sign, lab test, medical procedure, medication, clinical notes | - | - | - |

This table provides a subset of publicly available (or available upon request) longitudinal datasets in cancer research, with longitudinal data elements highlighted in red. These datasets typically encompass cancer screening, cancer treatment response, and large-scale biobanks. As some datasets are only accessible upon request, whether a data element is longitudinal was determined based on the available documentation.

[a]NLST = National Lung Screening Trial, PLCO = The Prostate, Lung, Colorectal and Ovarian, UKLWC = United Kingdom Collaborative Trial of Ovarian Cancer Screening Longitudinal Women's Cohort, CSAW-CC = Abbreviated Breast MRI and Digital Tomosynthesis Mammography in Screening Women With Dense Breasts, I-SPY = Investigation of Serial Studies to Predict Your Therapeutic Response with Imaging and moLecular Analysis, ReMIND = The Brain Resection Multimodal Imaging Database, PEDSnet = A National Pediatric Learning Health System, EA1141 = Abbreviated Breast MRI and Digital Tomosynthesis Mammography in Screening Women With Dense Breasts, QIN = Quantitative Imaging Network, MIMIC = The Medical Information Mart for Intensive Care. [b]GWAS = Genome-Wide Association Studies, WGS = Whole Genome Sequencing, WES = Whole exome sequencing, EWAS = epigenome-wide association, ctDNA = Circulating tumor DNA, SNP = Single Nucleotide Polymorphism, CNV = Copy Number Variation. [c]CT = Computed Tomography, FDG = Fludeoxyglucose, PET = Positron Emission Tomography, MRI = Magnetic Resonance Imaging, US = Ultrasound, DCE = Dynamic Contrast-enhanced, DBT = Digital Breast Tomosynthesis.

SUPPLEMENTAL TABLE II
FORMULAS IN LONGITUDINAL MODELING TECHNIQUES.

| Method | Equations |
|---|---|
| **Statistical Model** | |
| **Mixed Effect Model** | The general form of a linear mixed effects model is the following: $$Y_{ij} = \beta_{0i} + \beta_{1i}t_j + \varepsilon_{ij}$$ $$\beta_{0i}=\beta_0 + u_{0i}$$ $$\beta_{1i}=\beta_1 + u_{1i},$$ where $Y_{ij}$ is an outcome variable observed on individual $i$ at times $t_j$, $\beta_{0i}$ and $\beta_{1i}$ are individual-specific coefficients with both fixed ($\beta_0, \beta_1$) and random components ($u_{0i}, u_{1i}$), and $\varepsilon_{ij}$ is the residual error [17]. |
| **Parametric Empirical Bayes** | In the context of cancer screening, the parametric empirical bayes $Z$ score is calculated as: $$Z = \frac{Y - [\mu \times (1-\beta_n) + \bar{X}_i \times \beta_n]}{\sqrt{V(1-\beta_1\beta_n)}} \geq T,$$ where $Y$ is the current biomarker value, $\bar{X}_i$ is the mean of previous biomarker values for the individual, $\mu$ is the population mean of biomarker values, $V$ is the population variance of biomarker values, $\beta_1$ is the intraclass correlation coefficient within individuals over time, and $\beta_n$ is a weighted ICC depending on how many past values $n$ are available [18]. The screening result is considered positive whenever $Z$ is greater than a certain threshold $T$, which can be chosen based on the desired specificity. |
| **Dynamic Bayesian Network** | The state of each variable at time $t$ is conditionally dependent on a subset of variables from the current and previous time step. $$P(X_t \mid X_{t-1}) = \prod_i P(X_t^{(i)} \mid Pa(X_t^{(i)})),$$ where $X_t$ is the set of all state variables at timepoint $t$, $X_{t-1}$ is the set of all state variables at the previous timepoint, $X_t^{(i)}$ is the $i$-th variable at timepoint $t$, and $Pa(X_t^{(i)})$ indicates the parents of $X_t^{(i)}$, typically including variables from current and previous timepoint. |
| **Hidden Markov Model** | The probability of being in state $j$ after all observations equals the total of all potential outcome paths: $$p_t(j) = \sum_{i=1}^{N} p_{t-1}(i) T_{ij} E_j(o_t),$$ where $p_{t-1}(i)$ is the probability from the previous timestep in state $i$, $T_{ij}$ is the transition probability from state $i$ to $j$, $E_j(o_t)$ is the emission probability of having the observation at time $t$ given in state $j$. |
| **Deep Learning** | |
| **Recurrent Neural Network** | At each time step $t$, the network takes the input $x_t$ and the prior hidden state $h_{t-1}$ to compute the new hidden state $h_t$: $$h_t = W_1 x_t + W_2 h_{t-1},$$ where $W_1$ and $W_2$ represent weight matrices. |
| **Transformer** | Attention scores are calculated to assess how one timestamp $x_t$ attends to all other timestamps. They are computed using query ($Q$) and keys ($K$), which are linear projections of the time-series inputs $X$ with weights $W_Q$ and $W_K$. The attention scores are obtained by computing the dot product of $Q$ and $K$, where the result is scaled by the dimensionality of keys $d_k$ and then passed through the Softmax function for normalization. The final output is derived by taking the dot product of attention scores and value matrix V, which is the linear projection of X with weights $W_V$ [19]. $$Q = W_Q X; K = W_K X; V = W_V X$$ $$Attention(Q,K,V) = softmax\left(\frac{QK^T}{\sqrt{d_k}}\right)V$$ |
| **Self-supervised Learning** | Self-supervised learning typically employs deep learning architectures that utilize various loss functions without using labels. Here are some example loss functions that are generally used.<br><br>Mean Square Error (MSE) loss can be used to measure the difference between the true time difference ($x$) and the predicted time difference ($\hat{x}$). Also, it can be used in reconstruction to minimize the difference between input ($x$) and reconstructed output ($\hat{x}$). $$L_{MSE} = \lVert x - \hat{x} \rVert^2$$ Information Noise-Contrastive Estimation loss [20] can be used to pull paired embedding, $z_i$ and $z_j$, closer (datapoints from different timepoints of the same individual) and push the unpaired embedding, $z_i$ and $z_n$, apart. Cosine similarity is used to compute the distance between each pair of data. $\tau$ is a temperature parameter to control the softness of the probability distribution. $$L_{InfoNCE} = -\log \frac{\exp(sim(z_i, z_j)/\tau)}{\sum_{n=i}^{N} \exp(sim(z_i, z_n)/\tau)}$$ |
| **Reinforcement Learning** | Bellman equation is the fundamental concept in reinforcement learning. It is a recursive function which presents the value of being in the current state $s$, $V(s)$, using immediate reward ($R$) when action $a$ was taken and the value of the subsequent state $V(s')$. $$V(s) = \max_a \left(R(s,a) + \gamma V(s')\right)$$ |